\documentclass{aastex62}
\usepackage{amsmath} 
\submitjournal{ApJ}

\shorttitle{3-D isotropic feature of self-correlation level contours}
\shortauthors{Wu et al.}

\begin{document}

\title{3-D feature of self-correlation level contours at $10^{10}\  \mathrm{cm}$ scale in solar wind turbulence}

\correspondingauthor{Chuanyi Tu}
\email{chuanyitu@pku.edu.cn}

\author{Honghong Wu}
\affiliation{School of Earth and Space Sciences, Peking University, Beijing, People's Republic of China}

\author{Chuanyi Tu}
\affiliation{School of Earth and Space Sciences, Peking University, Beijing, People's Republic of China} 

\author{Xin Wang}
\affiliation{School of Space and Environment, Beihang University, Beijing, People's Republic of China} 

\author{Jiansen He}
\affiliation{School of Earth and Space Sciences, Peking University, Beijing, People's Republic of China} 

 \author{ Linghua Wang}
\affiliation{School of Earth and Space Sciences, Peking University, Beijing, People's Republic of China}



\begin{abstract}
The self-correlation level contours at $10^{10}\  \mathrm{cm}$ scale reveal a 2-D isotropic feature in both the slow solar wind fluctuations and the fast solar wind fluctuations. However, this 2-D  isotropic feature is obtained based on the assumption of axisymmetry with respect to the mean magnetic field. Whether the self-correlation level contours are still 3-D isotropic remains unknown. Here we perform for the first time a 3-D self-correlation level contours analysis on the solar wind turbulence. We construct a 3-D coordinate system based on the mean magnetic field direction and the maximum fluctuation direction identified by the minimum-variance analysis (MVA) method. We use data with 1-hour intervals observed by WIND spacecraft from 2005 to 2018. We find, on one hand, in the slow solar wind, the self-correlation level contour surfaces for both the magnetic field and the velocity field are almost spherical, which indicates a 3-D isotropic feature. On the other hand, there is a weak elongation in one of the perpendicular direction in the fast solar wind fluctuations. The 3-D feature of the self-correlation level contours surfaces cannot be explained by the existed theory.

\end{abstract}



\section{Introduction} \label{sec:intro}
    
The magnetohydrodynamic (MHD) turbulence exhibits anisotropic features as a result of the  preferred direction that the background magnetic field determines \citep{Shebalin1983JPP}. The solar wind is observed to be in a turbulence state \citep{Tu1995SSRv} and various studies proposed that  solar wind turbulence is 2-D anisotropic based on theories \citep{Oughton1994CUP, Goldreich1995ApJ}, simulations \citep{Cho2000ApJ} and observations related to the power spectral index \citep{Horbury2008PhRvL, Podesta2009ApJ, Chen2010PhRvL, Wicks2010MNRAS, He2013ApJ}, structure function \citep{Luo2010ApJL} and correlation function \citep{Matthaeus1990JGR, Dasso2005ApJ}. 

The “Maltese cross” is one 2-D pioneering work. It consists of two lobes, one elongated along to the mean field direction (slab-like fluctuations), and the other elongated along the perpendicular direction to the mean field direction (2-D fluctuations) and is obtained by a 2-D self-correlation analysis \citep{Matthaeus1990JGR}. \cite{Dasso2005ApJ} applied the correlation function method by using two-day-long data from Advanced Composition Explorer (ACE) spacecraft for the slow solar wind and the fast solar wind separately. They find that the fast wind mainly contains slab-like fluctuations and the slow wind 2-D fluctuations. However, \cite{Wang2019aApJ} find that the self-correlation function level contours of the magnetic field and the velocity field are 2-D isotropic for both the slow solar wind and the fast solar wind at $10^{10}\  \mathrm{cm}$ scale using the same method as \cite{Dasso2005ApJ} . 

The 2-D anisotropic studies have been extended to the 3-D scenario which includes not only the mean magnetic field direction, but also the perpendicular magnetic field fluctuation direction.  \cite{ Boldyrev2006PhRvL} predicted theoretically  that the solar wind turbulence is 3-D anisotropic with $l_{\mathrm{\parallel}}>l_{\mathrm{\perp 2}}>l_{\mathrm{\perp 1}}$, where $l_{\mathrm{\parallel}}$, $l_{\mathrm{\perp 2}}$, $l_{\mathrm{\perp 1}}$ are correlation lengths in the mean magnetic field direction, the perpendicular magnetic field fluctuation direction and the direction perpendicular to both, respectively. \cite{Chen2012ApJ} used the local structure function method to analyze the 3-D structure of turbulence in the fast solar wind in a new local coordinate system from the outer scale to the  proton gyroscale. \cite{Verdini2018ApJ} used the same local structure function method to analyze the 3-D structure taking the wind expansion effect into account.

In the present study, we perform the 3-D self-correlation function level contour analysis on the WIND spacecraft measurements. We construct the 3-D coordinate system using the mean magnetic field and the maximum variance fluctuation direction $L$ obtained by MVA method. In section \ref{sec:Data}, we describe the data and methods used in order to study the 3-D anisotropy, including the way to construct the 3-D coordinate system and get the 3-D contour surfaces. We show our observational results in section \ref{sec:Results}. In section \ref{sec:discussion}, we discuss our results and present our conclusions.

\section{Data and Method} \label{sec:Data}

We use data from the Wind spacecraft during 14 years from 2005 to 2018, when the spacecraft hovers at the Lagrangian point $L1$ in the solar wind. The magnetic field investigation \citep{Lepping1995SSRv} provides $3$ s resolution magnetic field data, and, the three-dimensional plasma analyzer \citep{Lin1995SSRv} measures the plasma data with a same cadence of $\Delta=3$ s. We cut the data set into 1-hour intervals with no overlap and require that the data gap accounts for less then $5\%$ in each interval. We remove the intervals with $max[|\delta B_j|]<2   \mathrm{nT}$, $ \ max[|\delta V_j|]<20 \mathrm{km}$, where $j$ indicates $x$, $y$, $z$ axis in the geocentric-solar-ecliptic (GSE) coordinate system, and $\delta$ means the variation between every $3$ s, in order to avoid the influence of shear magnetic field and shear flows.

For each interval $i$, we define the fluctuation as $\delta  \vec{U} =  \vec{U}- \bar{U}$, where $\vec{U}$ is either magnetic field $ \vec{B}$ or velocity $ \vec{V}$, and $ \bar{U}$ is obtained by performing a linear fit to $ \vec{U}$. The two-time-point self-correlation function of $\delta  \vec{U}$ is calculated as 
\begin{equation}
R_{U}(i,\tau)=<\delta  \vec{U}(t)\cdot\delta \vec{U}(t+\tau)>,
\end{equation}    
here, $\tau=0, \Delta,2\Delta,...,400\Delta$ is the time lag, and, $<>$ denotes an ensamble time average. In order to easily make camparison, we use the zero time lag self-correlation $R(i, 0)$ to normalize the self-correlation function and obtain $R_{uu}(i,\tau)=R_{U}(i,\tau)/R(i, 0)$. In this way, the $R_{uu}(i, \tau)$ at $\tau=0$ is always equal to $1$. According to the Taylor hypothesis \citep{Taylor1938RS}, we transfer the time lag to spatial lag using $r=\tau V_{\mathrm{SW}}$, where $r$ is the spatial lag and $V_{\mathrm{SW}}$ is the mean flow velocity in the corresponding interval $i$. 

\begin{figure}[ht!]
\includegraphics{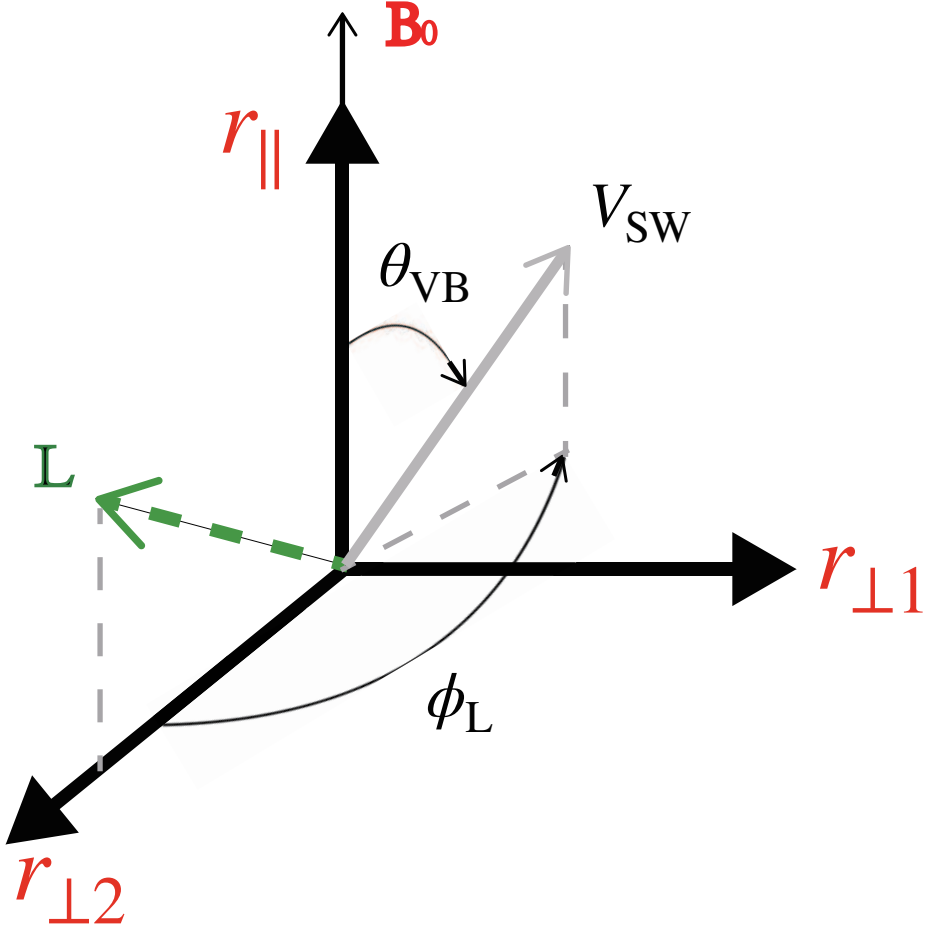}
\caption{3-D reference coordinate system used to compute correlation functions. For each 1-hour interval,  $r_{\mathrm{\parallel}}$ corresponds to the direction of the mean magnetic field $\vec{B}_\mathrm{0}$, and, the projection of the maximum fluctuation direction $L$ on the perpendicular plane is defined as $r_{\mathrm{\perp 2}}$, and, $r_{\mathrm{\perp 1}}$ completes this orthogonal coordinate system. $\theta_{\mathrm{VB}}$ is the angle between the mean magnetic field and the solar wind velocity, and, $\phi_{\mathrm{L}}$ is the angle between $r_{\mathrm{\perp 2}}$ and the projection of the solar wind velocity on the plane perpendicular to $\vec{B}_\mathrm{0}$.}\label{fig:figure1}
\end{figure}

\cite{Wang2019aApJ} has shown the isotropic feature of the self-correlation level contours in a 2-D coordinate system. We extend this system to 3-D by introducing the maximum variance direction $L$, which is determined by performing minimum-variance analysis (MVA) method \citep{Sonnerup1967JGR} to the magnetic field data. This 3-D coordinate system uses the mean magnetic field $\vec{B}_\mathrm{0}$ and the projection of maximum variance direction $L$ in the plane perpendicular to the mean field as $r_{\mathrm{\parallel}}$ and $r_{\mathrm{\perp 2}}$ components, respectively. $r_{\mathrm{\perp 1}}$ components completes this orthogonal coordinate system. Any angles greater than $90^\circ$ are reflected below  $90^\circ$. In Figure \ref{fig:figure1}, we show the angle $\theta_{\mathrm{VB}}$ between the directions of $V_{\mathrm{SW}}$ and $\vec{B}_\mathrm{0}$ and the angle $\phi_{\mathrm{L}}$ between  $r_{\mathrm{\perp 2}}$ direction and the component of $V_{\mathrm{SW}}$ perpendicular to $\vec{B}_\mathrm{0}$ for each interval $i$.

 \begin{figure}[ht!]
\includegraphics[width=\linewidth]{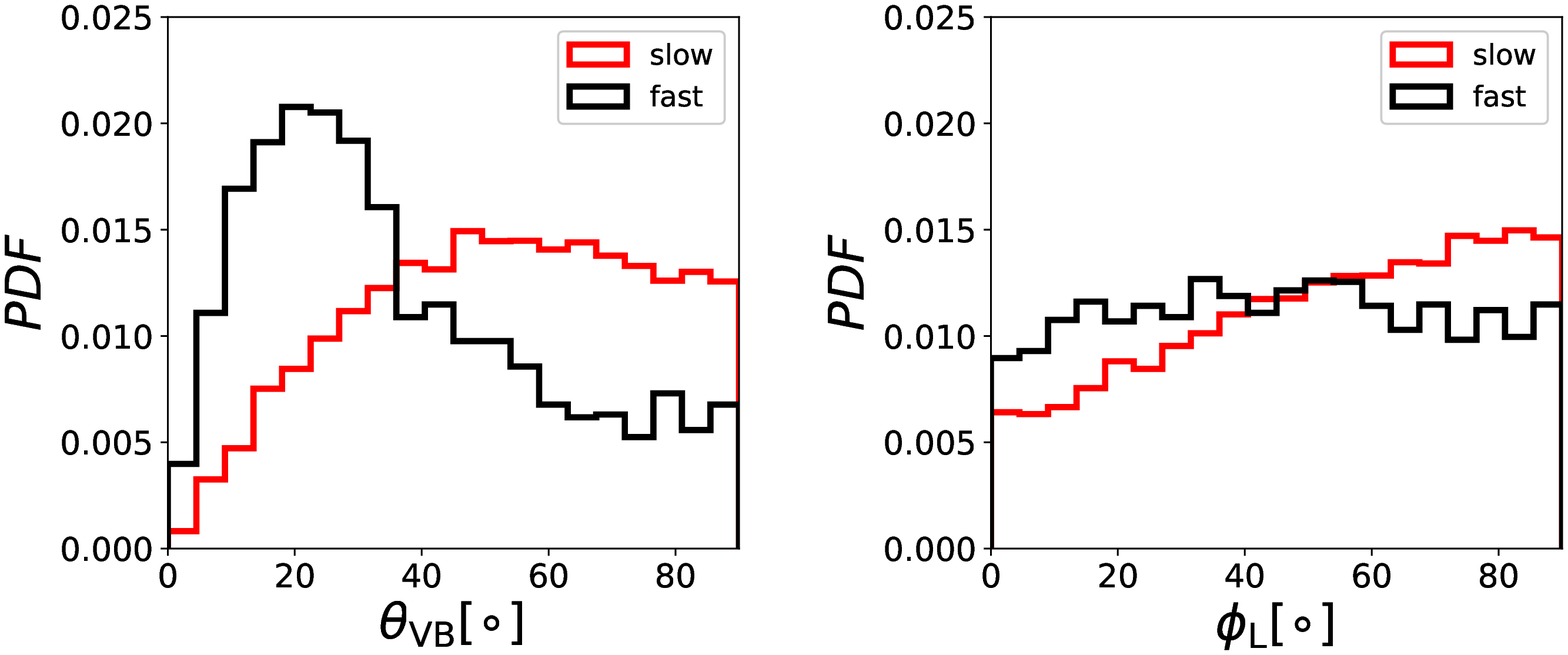}
\caption{Probability density function of $\theta_{\mathrm{VB}}$ (left) and $\phi_{\mathrm{L}}$ (right). The red and black histograms are for the slow wind and the fast wind, respectively. }\label{fig:figure2}
\end{figure} 

We find $23083$ intervals in the slow solar wind ($V_{\mathrm{SW}}<400$ km/s ) and $3347$ intervals in the fast solar wind ($V_{\mathrm{SW}}>500$ km/s ) and study their 3-D self-correlation level contours separately. The probability density function of $\theta_{\mathrm{VB}}$ in the left panel of Figure \ref{fig:figure2} shows that the magnetic field is more oblique to the solar wind velocity in the slow wind than in the fast wind, which is consistent with the Parker Spiral theory. In the right panel, we show in the slow wind, there are more intervals with perpendicular $\phi_{\mathrm{L}}$ than parallel $\phi_{\mathrm{L}}$, while the fast wind group has a roughly even distribution over $0^\circ$ and $90^\circ$.

For each group, we bin  $\theta_{\mathrm{VB}}$ and $\phi_{\mathrm{L}}$ into $15^\circ$ bins and calculate the average of the normalized spatial self-correlation functions as follows:
\begin{equation}
R_{uu}(\theta_{\mathrm{VB}}^m,\phi_{\mathrm{L}}^n,r)=\frac{1}{n(\theta_{\mathrm{VB}}^m,\phi_{\mathrm{L}}^n)} \sum_{\substack{\theta_{\mathrm{VB}}^m-7.5<=\theta_{\mathrm{VB}}(i)<\theta_{\mathrm{VB}}^m+7.5,\\ \phi_{\mathrm{L}}^n-7.5<=\phi_{\mathrm{L}}(i)<\phi_{\mathrm{L}}^n+7.5}}R_{uu}(i,r)
\end{equation}  
where $n(\theta_{\mathrm{VB}}^m,\phi_{\mathrm{L}}^n)$ is the number of the intervals in corresponding bin, and, $\theta_{\mathrm{VB}}^m=15^\circ m+7.5^\circ;\phi_{\mathrm{L}}^n=15^\circ n+7.5^\circ; m,n=0,1,2,...,5.$

We obtain 36 averaged self-correlation functions for 36 $(\theta_{\mathrm{VB}},\phi_{\mathrm{L}})=15^\circ \times 15^\circ$ bins. We analyze the contours at level $R_{uu}(\theta_{\mathrm{VB}},\phi_{\mathrm{L}},r)=1/e \approx 0.368$, and obtain a $r_{\mathrm{level}}$ value by linear interpolation for each $(\theta_{\mathrm{VB}},\phi_{\mathrm{L}})$. In order to plot the contour surface in the 3-D coordinate system, we transform $(\theta_{\mathrm{VB}},\phi_{\mathrm{L}},r_{\mathrm{level}})$ into $(r_{\mathrm{\perp 1},} r_{\mathrm{\perp 2}}, r_{\mathrm{\parallel}})$ by using $ r_{\perp1}=r_{\mathrm{level}}\sin\theta_{\mathrm{VB}}\sin\phi_{\mathrm{L}}$, $r_{\mathrm{\perp 2}}=r_{\mathrm{level}}\sin\theta_{\mathrm{VB}}\cos\phi_{\mathrm{L}}$, $r_{\mathrm{\parallel}}=r_{\mathrm{level}}\cos\theta_{\mathrm{VB}}$. We reflect the surface in the first octant into the other seven octants based on the assumption of reflectional symmetry. The result is presented in the next section.
 
\section{Results} \label{sec:Results}

\begin{figure}[ht!]
\includegraphics[width=\linewidth]{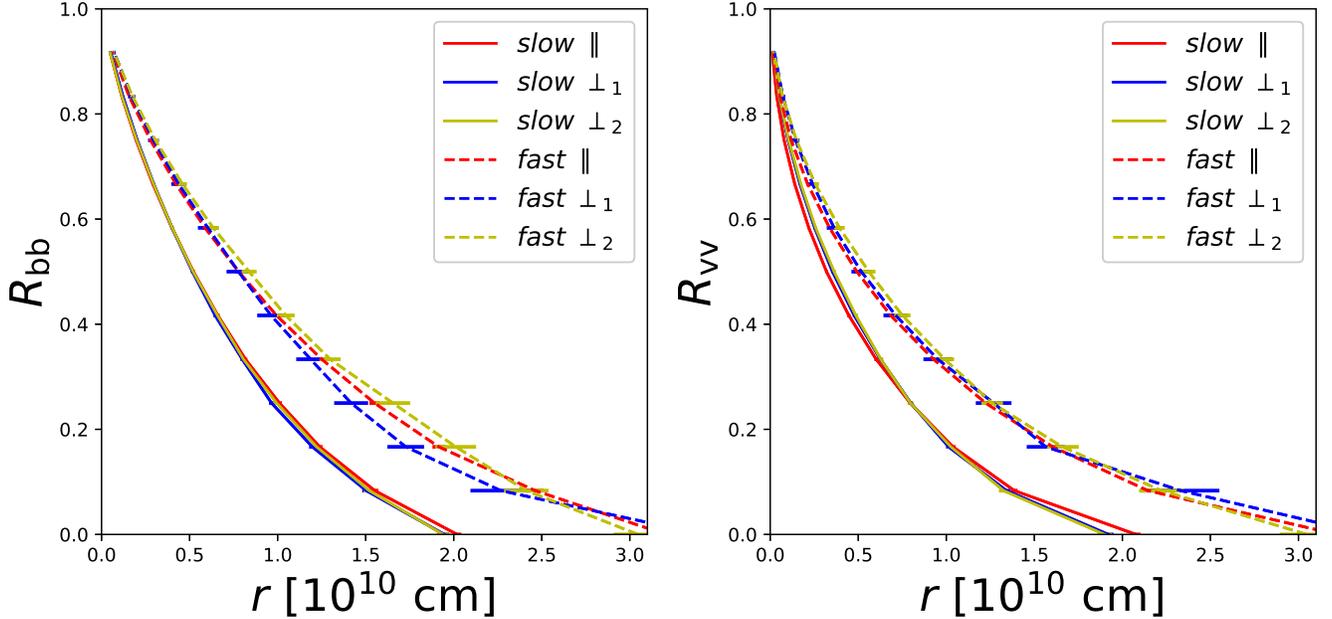}
\caption{Left panel: Averaged normalized self-correlation functions $R_{\mathrm{bb}}(r) $ of 1-hour-long magnetic field data. The solid and dashed lines are for the slow wind and the fast wind. Red, blue,  and yellow colors correspond the $r_{\mathrm{\parallel}}$, $r_{\mathrm{\perp 1}}$, and $r_{\mathrm{\perp 2}}$ directions, respectively. The error bar shows the standard error of $r_{\mathrm{level}}$ for a given $R_{\mathrm{bb}}$. Right panel:  Averaged normalized self-correlation functions $R_{\mathrm{vv}}(r) $ of 1-hour-long velocity data, in the same manner as the left panel. }\label{fig:figure3}
\end{figure}

Figure \ref{fig:figure3} shows the averaged self-correlation functions in $r_{\mathrm{\perp 1}}$, $r_{\mathrm{\perp 2}}$, and $r_{\mathrm{\parallel}}$ directions, which correspond to the following angular bins:

\begin{align}
r_{\mathrm{\perp 1} }\rightarrow \ (75^\circ<=\theta_{\mathrm{VB}}<=90^\circ,\ 75^\circ<=\phi_{\mathrm{L}}<=90^\circ),\\
r_{\mathrm{\perp 2}} \rightarrow \ (75^\circ<=\theta_{\mathrm{VB}}<=90^\circ,\ 0^\circ<=\phi_{\mathrm{L}}<15^\circ),\\
r_{\mathrm{\parallel}}\rightarrow \ (0^\circ<=\theta_{\mathrm{VB}}<15^\circ,\ 0^\circ<=\phi_{\mathrm{L}}<=90^\circ).
\end{align}
In the left panel of Figure \ref{fig:figure3}, we present the averaged magnetic self-correlation functions with standard error bars for both the slow solar wind (solid lines) and the fast solar wind (dashed lines). It is hard to distinguish the functions of the three directions for the slow wind. The phenomenon of the functions almost overlapping with each other means the 3-D isotropic feature of the self-correlation level contours in slow solar wind turbulence. For the fast wind, there is a slightly elongation along the $r_{\mathrm{\perp 2}}$ direction in the perpendicular plane. Note that the magnetic self-correlation function of the fast wind is larger than that of the slow wind. When we consider self-correlation function with respect to the time lag instead of the spatial lag, the magnetic self-correlation functions for both the slow wind and the fast wind are almost the same (not shown). In the right panel, we show the averaged velocity self-correlation functions. They have almost the same features as the averaged magnetic self-correlation functions except there is no clear elongation along the $r_{\mathrm{\perp 2}}$ direction.


\begin{figure}[ht!]
\includegraphics[width=\linewidth]{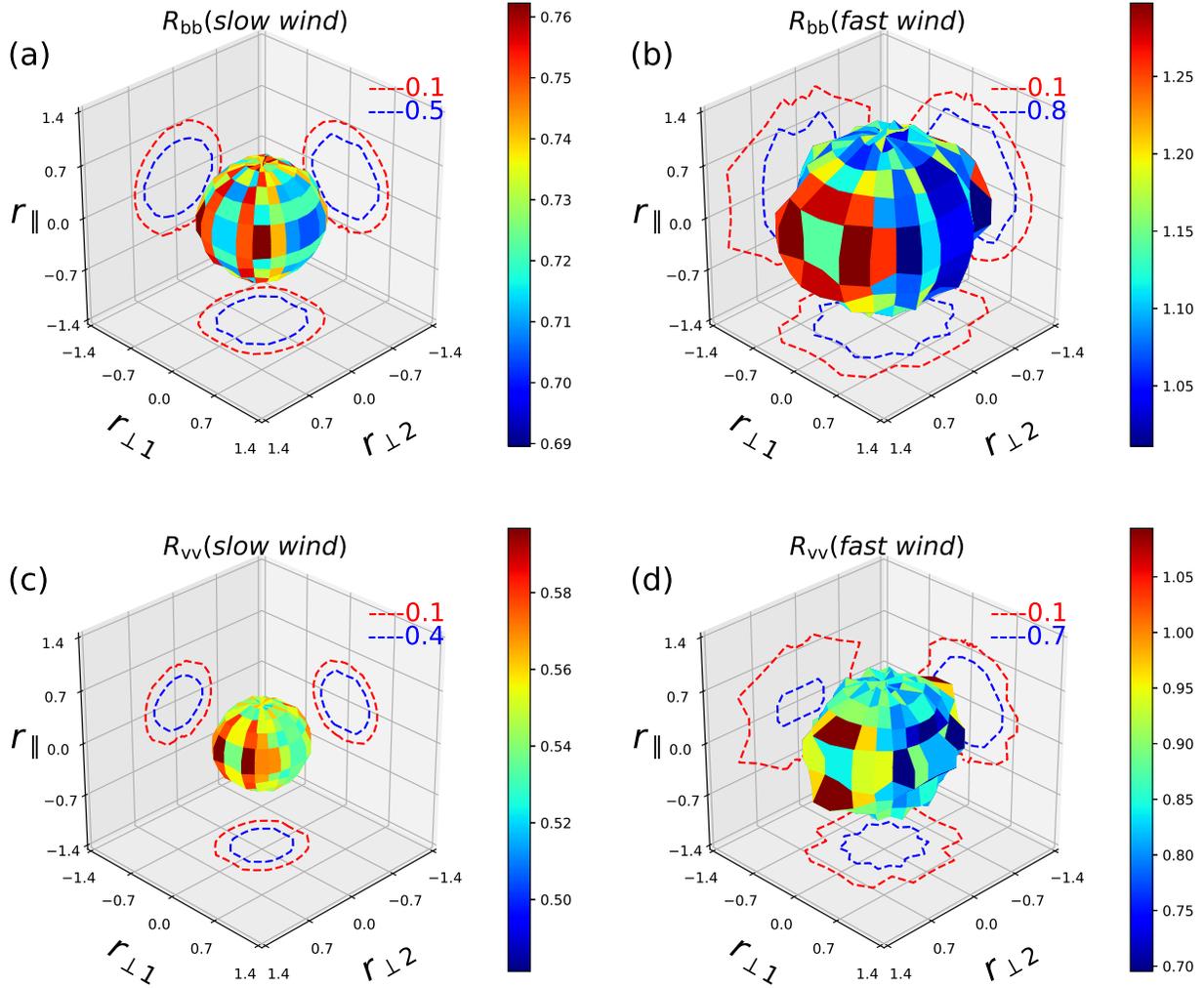}
\caption{ 3-D self-correlation level contour surface at level $R_{uu}=0.368$ of (a) magnetic field in the slow wind; (b) magnetic field in the fast wind; (c) velocity field in the slow wind; (d) velocity field in the fast wind. The color represents $r_{\mathrm{level}}\  [10^{10}\ \mathrm{cm}]$, which is the distances from the origin. The dashed red and blue lines in $r_{\mathrm{\perp 1}}=-1.4$ plane are projections of the intersection lines of the surface with two planes $r_{\mathrm{\perp 1}}=A1$ and  $r_{\mathrm{\perp 1}}=A2$, respectively, where $A1$ and $A2$ are shown in the legends with the corresponding colors in the corresponding panel ; the dashed red and blue lines in $r_{\mathrm{\perp 2}}=-1.4$ plane are projections of the intersection lines of the surface with two plane $r_{\mathrm{\perp 2}}=A1$ and  $r_{\mathrm{\perp 2}}=A2$, respectively; the dashed red and blue lines in $r_{\mathrm{\parallel}}=-1.4$ plane are projections of the intersection lines of the surface with two plane $r_{\mathrm{\parallel}}=A1$ and  $r_{\mathrm{\parallel}}=A2$, respectively. }\label{fig:figure4}
\end{figure}

We show the 3-D self-correlation level contour surfaces at level $R_{uu}=0.368$ in Figure \ref{fig:figure4}. In Figure \ref{fig:figure4}(a), the slow wind magnetic field self-correlation function contour surface is almost a spherical surface. The projection closed curves on the 2-D plane are plotted to help visualize the isotropic feature. They are almost round and almost identical to each other, which confirm the isotropic result. In Figure \ref{fig:figure4}(b), the fast wind magnetic field self-correlation level contour surface departs a bit from a spherical surface. $r_{\mathrm{level}}$ in the $r_{\mathrm{\perp 2}}$ direction is slightly longer. The projection closed curves on three planes are not round and have different sizes between each other. In Figure \ref{fig:figure4}(c), the slow wind velocity field self-correlation level contour surface is almost spherical and the projection closed curves on the 2-D planes is also round and identical to each other, which shows a clear isotropic feature as the magnetic field. In Figure \ref{fig:figure4}(d), the fast wind velocity field self-correlation level contour surface has a similar shape with that of magnetic field. The similarity between the magnetic field and velocity field contour shape supports the applicability of the data analysis technique here. We should also note that, the $r_{\mathrm{level}}$ is shorter for the slow wind than for the fast wind and shorter for the velocity field than for the magnetic field.

\begin{figure}[ht!]
\includegraphics[width=\linewidth]{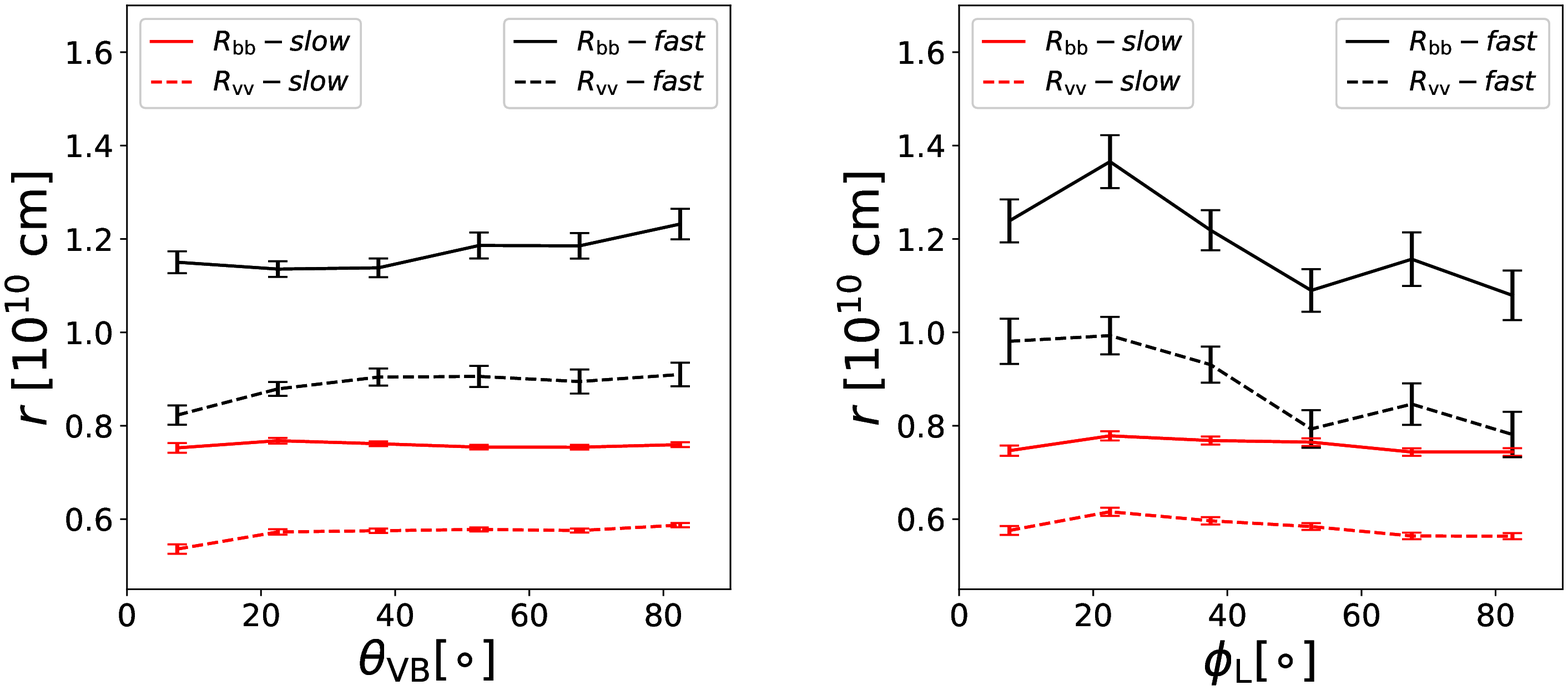}
\caption{Left panel: averaged $r_{\mathrm{level}}$ in each $\theta_{\mathrm{VB}}$ bin. The solid and dashed lines are for the magnetic field and the velocity field. And, the red and black lines indicate the slow wind and the fast wind, respectively. The error bars show the standard errors of the averaged $r_{\mathrm{level}}$. Right panel:  averaged $r_{\mathrm{level}}$ in each $\phi_{\mathrm{L}}$ bin, in the same manner as the left panel.  }\label{fig:figure5}
\end{figure}

In order to evaluate the unevenness shown in Figure \ref{fig:figure4}, we reduce the 3-D surface into the line trend with $\theta_{\mathrm{VB}}$ and $\phi_{\mathrm{L}}$ , as shown in Figure \ref{fig:figure5}. We calculate the averaged $r_{\mathrm{level}}$ in 6 $\theta_{\mathrm{VB}}$ bins from $0^\circ<=\phi_{\mathrm{L}}<=90^\circ$ with a weight of interval number in each $\phi_{\mathrm{L}}$ bin. The result is shown in the left panel. We can clearly see that the variation with $\theta_{\mathrm{VB}}$ is rather small for both the slow wind and the fast wind and for both the magnetic field and the velocity field. We calculate the average $r_{\mathrm{level}}$ in 6 $\phi_{\mathrm{L}}$ bins from $60^\circ<=\theta_{\mathrm{VB}}<=90^\circ$ with a weight of interval number in the two $\theta_{\mathrm{VB}}$ bins. The result is shown in the right panel. For the slow wind, the variation with $\phi_{\mathrm{L}}$ is very small; while for the fast wind, there is a weak elongation along $r_{\mathrm{\perp 2}}$. Again, it is easily seen that $r_{\mathrm{level}}$ is shorter for the slow wind and the velocity field. 

\begin{figure}[ht!]
\includegraphics[width=\linewidth]{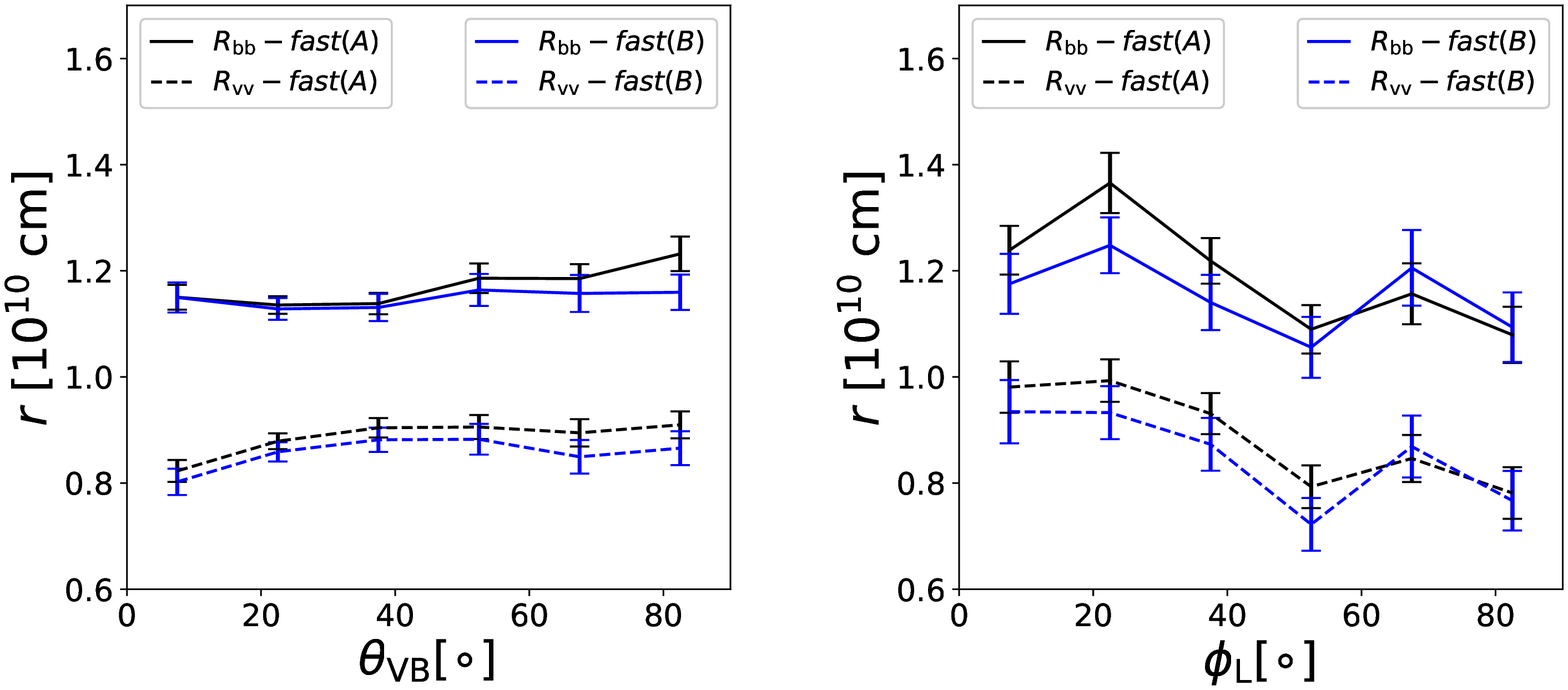}
\caption{Left panel: averaged $r_{\mathrm{level}}$ in each $\theta_{\mathrm{VB}}$ bin. The solid and dashed lines are for the magnetic field data and velocity data. And, the black and blue lines indicate the fast wind group A and the fast wind group B, respectively. The error bars show the standard errors of the averaged $r_{\mathrm{level}}$. Right panel:  averaged $r_{\mathrm{level}}$ in each $\phi_{\mathrm{L}}$ bin, in the same manner as the left panel.  }\label{fig:figure6}
\end{figure}

In Figure \ref{fig:figure6}, we show the variations with $\theta_{\mathrm{VB}}$ and $\phi_{\mathrm{L}}$ for the fast solar wind. The black solid and black dashed lines are the same as in  Figure \ref{fig:figure5}. We check the data intervals in the fast wind (group A) and further rule out the intervals with large gradient by visual inspection. We reserve $2272$ cases (group B). The variations with $\theta_{\mathrm{VB}}$ and $\phi_{\mathrm{L}}$ for this new fast wind group B are calculated and shown in blue lines in Figure \ref{fig:figure6}. The anisotropy for the fast group becomes weaker after we remove the structures more strictly.
 
\section{Discussion and  Conclusions} \label{sec:discussion}

We present for the first time the 3-D self-correlation level contours of the magnetic field and the velocity field at $10^{10} \ \mathrm{cm}$ scale based on WIND spacecraft measurements from 2005 to 2018. We construct a 3-D coordinate system according to the mean magnetic field direction and the maximum variance direction $L$ of the magnetic field. The self-correlation contour surfaces at level $R_{uu}\approx 1/e$  in the slow solar wind are 3-D isotropic for both the magnetic field and the velocity field. The self-correlation contour surfaces at level $R_{uu}\approx 1/e$  in the fast solar wind show weak anisotropic feature in the perpendicular plane with an elongation along $r_{\mathrm{\perp 2}}$. However, the anisotropy becomes weaker when we exclude the intervals with structures more strictly. 

The 3-D coordinate system constructed here is consistent with the 3-D coordinate system presented by \citep{Chen2012ApJ} if we consider the maximum variance direction $L$  as the $(\vec{B_\mathrm{1}}-\vec{B_\mathrm{2}})$ in their work.  \cite{Chen2012ApJ} present a structure function analysis in a scale-dependent 3-D coordinate system defined as follows: for each pair of points, the local mean field $B_{\mathrm{local}}=(\vec{B_\mathrm{1}}+\vec{B_\mathrm{2}})/2$ was calculated as one axis and the local perpendicular fluctuation direction $B_{\mathrm{local}}\times [(\vec{B_\mathrm{1}}-\vec{B_\mathrm{2}})\times B_{\mathrm{local}}]$ as another axis. \cite{ Verdini2018ApJ} perform a structure function analysis in the same 3-D coordinate system. However, our maximum variance direction $L$ is based on the whole interval, while their local perpendicular fluctuation direction is adjusted for every two time instances.

The surfaces of the fast wind have an approximately $1.5$ times larger spatial size than the surfaces of the slow wind for both the magnetic field and the velocity field. This result is consistent with the result of \cite{Wang2019aApJ}. However, we also find that the correlation level contours are similar for the fast and slow wind without transferring the time lag into the spatial lag through the Taylor hypothesis. The reason why the fast wind has larger correlation level contour surfaces than the slow wind is unknown. The surfaces  of the magnetic field have an approximately $1.3$ times larger spatial size than the surfaces of the velocity field for both the fast wind and the slow wind. This result provides constraints for the solar wind turbulence theory.

The scale of $10^{10} \ \mathrm{cm}$ corresponds more or less to the low-frequency break scale universally observed in the fast solar wind turbulence. The 3D quasi-isotropic feature of the self-correlation level contours in the fast solar wind is consistent with Figure 1 by \cite{Wicks2010MNRAS} that the power spectrum is isotropic at the low-frequency break. Whether the anisotropy of the 3-D self-correlation level contours increase at smaller and smaller scales needs further study. That the contour of the velocity field is similar to the contour of the magnetic field is reasonable since magnetic and velocity fluctuations are coupled. Currently, we have no idea why the size of the magnetic contour is larger. 

Recently, \cite{Bruno2019aa} find that the low-frequency break is also present in the slow solar wind magnetic spectra. They show a case with the break located around $10^{-4} \ \mathrm{Hz}$ and an average velocity of $316  \ \mathrm{km/s}$. Our scale in the slow solar wind is smaller than the approximate break scale $3\times 10^{11} \ \mathrm{cm}$ and is in the typical quasi-Kolmogorov range, as in \cite{Bruno2019aa}. Our 3-D self-correlation level contour analysis of both the magnetic and velocity field shows a 3-D isotropic feature. We consider the self-correlation level contours represent the angular feature of the turbulence eddies. These results are not consistent with the predictions of the existed MHD theories. The 3-D isotropic feature of self-correlation level contour supports the Kolmogorov's theory \citep{Kolmogorov1941ANSD}. How to interpret this result in the slow solar wind requires further investigation. 
\acknowledgments

We thank the CDAWEB for access to the Wind data, Dr. Liping Yang and Dr. Junxiang Hu for helpful discussions. This work at Peking University and Beihang University is supported by the National Natural Science Foundation of China under contract Nos. 41474147, 41504130, 41874199, and 41674171.

\end{document}